\documentclass[conference,screen]{IEEEtran} 
\usepackage{cite}
\usepackage{xspace} 
\usepackage{amsmath,amssymb,amsfonts}
\usepackage{algorithmic}
\usepackage{graphicx}
\usepackage{textcomp}
\usepackage{xcolor}
\usepackage{stfloats}
\usepackage{balance}
\usepackage{hyperref}
\usepackage{soul}
\usepackage{enumitem}
\usepackage{booktabs}
\usepackage{caption}
\usepackage{siunitx}
\usepackage{tabularx}
\usepackage{subcaption}
\usepackage{tcolorbox}
\usepackage{array}
\usepackage[retain-zero-uncertainty=true]{siunitx}
\tcbuselibrary{skins}

\def\BibTeX{{\rm B\kern-.05em{\sc i\kern-.025em b}\kern-.08em
    T\kern-.1667em\lower.7ex\hbox{E}\kern-.125emX}}

\makeatletter
\newcommand{\linebreakand}{%
  \end{@IEEEauthorhalign}
  \hfill\mbox{}\par
  \mbox{}\hfill\begin{@IEEEauthorhalign}
}
\makeatother

\newcommand\instruct{Instruction-editing\xspace}
\newcommand\inpaint{Inpainting\xspace}
\newcommand\refining{Inpainting with Refinement\xspace}
\newcommand\ours{Knowledge Distillation\xspace}

\newcommand{\davetwo}{\mbox{DAVE-2}\xspace}
\newcommand{\head}[1]{\noindent\textbf{#1.}}

\def\BibTeX{{\rm B\kern-.05em{\sc i\kern-.025em b}\kern-.08em
    T\kern-.1667em\lower.7ex\hbox{E}\kern-.125emX}}

\begin{document}

\pagenumbering{arabic} 
\pagestyle{plain}

\title{
Efficient Domain Augmentation for Autonomous Driving Testing Using Diffusion Models\\
\thanks{This research was partially supported by project EMELIOT, funded by MUR under the PRIN 2020 program (n. 2020W3A5FY), by the Bavarian Ministry of Economic Affairs, Regional Development and Energy, by the TUM Global Incentive Fund, and by the EU Project Sec4AI4Sec (n. 101120393).}
}

\author{\IEEEauthorblockN{Luciano Baresi\textsuperscript{1}, Davide Yi Xian Hu\textsuperscript{1}, Andrea Stocco\textsuperscript{2,3}, Paolo Tonella\textsuperscript{4}}\IEEEauthorblockA{\textsuperscript{1}\textit{Politecnico di Milano}, Milano, Italy, {luciano.baresi, davideyi.hu}@polimi.it}\ \IEEEauthorblockA{\textsuperscript{2}\textit{Technical University of Munich}, Munich, Germany, \ andrea.stocco@tum.de}\ \IEEEauthorblockA{\textsuperscript{3}\textit{fortiss GmbH}, Munich, Germany, \ stocco@fortiss.org}\ \IEEEauthorblockA{\textsuperscript{4}\textit{Software Institute - USI}, Lugano, Switzerland, \ paolo.tonella@usi.ch}}

\IEEEoverridecommandlockouts
\IEEEpubid{\makebox[\columnwidth]{XXX/\$31.00~\copyright2024 IEEE \hfill} \hspace{\columnsep}\makebox[\columnwidth]{ }}


\IEEEpubidadjcol

\maketitle

\begin{abstract}
Simulation-based testing is widely used to assess the reliability of Autonomous Driving Systems (ADS), but its effectiveness is limited by the operational design domain (ODD) conditions available in such simulators. To address this limitation, in this work, we explore the integration of generative artificial intelligence techniques with physics-based simulators to enhance ADS system-level testing.
Our study evaluates the effectiveness and computational overhead of three generative strategies based on diffusion models, namely instruction-editing, inpainting, and inpainting with refinement.
Specifically, we assess these techniques' capabilities to produce augmented simulator-generated images of driving scenarios representing new ODDs. We employ a novel automated detector for invalid inputs based on semantic segmentation to ensure semantic preservation and realism of the neural generated images.
We then performed system-level testing to evaluate the ability of the ADS to generalize to newly synthesized ODDs.
Our findings show that diffusion models help to increase the coverage of ODD for system-level ADS testing. Our automated semantic validator achieved a percentage of false positives as low as 3\%, retaining the correctness and quality of the images generated for testing. Our approach successfully identified new ADS system failures before real-world testing. 
\end{abstract}

\begin{IEEEkeywords}
autonomous driving systems, deep learning testing, diffusion models, generative AI
\end{IEEEkeywords}

\section{Introduction}\label{sec:introduction}

Before deploying Autonomous Driving Systems (ADS) on public roads, extensive simulation-based testing~\cite{grigorescu2020survey} is performed to ensure that these systems can effectively handle the scenarios of the Operational Design Domain (ODD), which defines the specific conditions under which an ADS is designed to operate~\cite{ISO34504,DBLP:journals/tiv/HuLHTHC24}.
Current driving simulators are based on game engines (e.g., Unity~\cite{unity} or Unreal~\cite{unreal}) and enable testing using a predefined set of ODD conditions.
Simulators are often limited in the range of ODD they represent, as they primarily focus on photorealistic rendering and accurate physics representation. Thus, they fail to cover many ODD scenarios that are instead critical for testing ADS. 
This limitation hinders the effectiveness of ADS system-level testing~\cite{Gambi:2019:ATS:3293882.3330566,2020-Riccio-FSE}, particularly for edge cases \textit{beyond} the predefined ODD conditions available in these simulators. 

Enhancing ODD coverage in a simulator typically requires developing new conditions within the simulator engine, a task that demands significant domain knowledge and development effort. Moreover, even if improving the simulation platform were feasible, recent research highlighted the problem of fidelity gap between the virtual environments represented in the simulators and the real world~\cite{2022-Stocco-TSE}.

Generative Artificial Intelligence (GenAI) solutions have been employed to improve and extend the range of ODD conditions for ADS testing~\cite{DBLP:conf/kbse/ZhangZZ0K18, DBLP:conf/qrs/PanAF21,DBLP:conf/icml/LiPZL21,deeproad,2022-Stocco-TSE,biagiola2023better}. However, approaches such as DeepRoad~\cite{deeproad} and TACTICS~\cite{DBLP:conf/icml/LiPZL21} use GenAI techniques that require mining a training corpus of ODD data and focus on offline model-level testing of individual Deep Neural Networks (DNNs) against different neural-generated images. Furthermore, these approaches have not been evaluated for system-level testing, which is crucial for assessing the safety requirements of autonomous driving~\cite{DBLP:journals/ese/StoccoPT23}.
On the other hand, closed-loop data-driven approaches, such as DriveGAN~\cite{DBLP:conf/cvpr/KimP0F21}, use GenAI to produce a continuous stream of driving images. However, their primary limitations lie in their dependence on learned physics models, which may be inaccurate or unrealistic. Consequently, these approaches are mainly useful for training data augmentation, rather than for testing. The lack of robust physics engines leads to inconsistent physical behaviors and interactions, making it impossible to simulate system-level failures such as collisions or vehicles driving off-road, as such scenarios are not available in the training data.

In this paper, we evaluate three different augmentation strategies based on state-of-the-art pre-trained diffusion models~\cite{DBLP:conf/cvpr/BrooksHE23,DBLP:conf/iclr/MengHSSWZE22,DBLP:conf/cvpr/RombachBLEO22}: \instruct, \inpaint, and \refining, used to expand the set of ODD conditions in a driving simulator via real-time image-to-image translation. Unlike existing GenAI techniques that require explicit training, diffusion models only require an input image and conditioning inputs that represent the transformation to be applied (e.g., a textual description).

As the diffusion models' output can be affected by artifacts, distortions, or inconsistencies that undermine the effectiveness of ADS testing~\cite{2024-Lambertenghi-ICST}, our methodology includes an automated validation technique based on image semantics and segmentation maps. This check is performed to assess the correctness and reliability of the generated images (e.g., ensuring that the original road shape is preserved after the augmentation). 

Although diffusion models can generate diverse images for testing, their direct use in a simulator faces two main challenges. First, diffusion models exhibit high inference times, in the order of seconds per image, which makes them impractical for real-time applications. Second, simulation platforms typically generate multiple image frames per second, but diffusion models lack the rendering consistency required for a coherent simulation, resulting in consecutive frames that may be drastically different.
To address these limitations, our approach leverages knowledge distillation~\cite{DBLP:journals/corr/HintonVD15} by integrating diffusion models with a cycle-consistent network~\cite{DBLP:conf/iccv/ZhuPIE17} to ensure domain generation consistency and high throughput.

In our experiments, our approach generated images that exhibited a validity rate between $52\%$ and $99\%$, the best approach being \inpaint. When used as a rendering engine within the Udacity simulator to produce $108$ simulations using various ODD conditions, our approach was able to reveal a total of $600$ failures across four lane-keeping ADS, $20$ times more than using the ODD conditions available in the simulator, at the cost of an increase in simulation time of $2\%$. Furthermore, we evaluated the generalizability of our approach to complex urban scenarios using the CARLA simulator and InterFuser~\cite{DBLP:conf/corl/ShaoW00022}, successfully exposing 25 red-light infractions and 3 vehicle collisions previously undetected.

Our paper makes the following contributions.

\noindent
\head{Approach} A system-level testing technique for ADS that combines different GenAI (e.g., diffusion models and cycle-consistent generative networks) as a rendering engine and physics-based simulators for effective failure detection. 
This novel combination of techniques achieves high ODD diversity, realism, semantic preservation, temporal consistency, and high throughput, while the simulator's underlying physics ensures accurate representation.
Our approach is integrated in the Udacity simulator for self-driving cars and is publicly available~\cite{replication-package}. To the best of our knowledge, this is the first solution that uses GenAI techniques within a driving simulator to improve the ODD coverage of DNN-based ADS. 

\noindent
\head{Evaluation} An empirical study on the validity and realism of neural-generated driving images and their use for system-level ADS testing.

\noindent
\head{Dataset} A dataset of more than $1$ million pairs of images and $52$ OOD conditions based on the Udacity simulator and a dataset of $1$ million images and $9$ OOD conditions based on the CARLA simulator. These datasets can be used to evaluate the generalizability of ADS to novel environmental conditions, as well as the performance of failure prediction systems.
\section{Background}\label{sec:background}

\subsection{System Level Testing of ADS}\label{sec:background:testing}

ADS must adhere to specific regulations that establish essential safety requirements for public acceptance and large-scale deployment.
In particular, standards such as the ISO/PAS 21448 Safety of the Intended Function (SOTIF) standard~\cite{sotif} or the UN Regulation No 157 (2021/389) concerning the approval of vehicles with regards to Automated Lane Keeping Systems~\cite{UNRegulation157}, demand extensive coverage of the Operational Design Domain (ODD) conditions. 
The ODD should describe the conditions under which the automated vehicle is intended to drive autonomously, such as roadway types; geographic area; speed range; environmental conditions (weather as well as day/night time); and other domain constraints. 
In this work, we focus on the \textit{ODD} conditions that \textit{visually} impact the environment and the DNNs of the ADS. Specifically, we used the conditions described in the standard ISO 34505~\cite{ISO34504} ``Scenery Elements (Section 9)'' and ``Environmental Conditions (Section 10)'', some examples being different geographic areas (e.g., European cities or coastal areas) and weather conditions (e.g., cloudy or rainy weather as well as day/night). 

The safe deployment of ADS requires a thorough exploration of ODDs through simulated and in-field testing. Due to significant time, space, and cost constraints associated with in-field testing (i.e., real-world testing with physical vehicles), simulation-based testing has become the standard option for system-level testing of ADS~\cite{2022-Stocco-TSE}. 
Driving simulators can generate data and conditions that closely mimic those encountered in real-world scenarios~\cite {DBLP:conf/icst/HaqSNB20}. 
To test the limits of the ADS, a simulator produces a vast amount of highly consistent data through synthetic input generation using a 3D image rendering engine. 
However, a comprehensive test dataset must not only be statistically significant in volume but also adequately represent the diverse ODD conditions. 
This is a major limitation of current driving simulators, which often have restricted ODD coverage, which is essential for comprehensive testing and fault exposure~\cite{grigorescu2020survey}. 

\subsection{Vision Generative AI}\label{sec:background:vision}

Generative AI has significantly advanced various vision tasks by enabling the creation of realistic and diverse data~\cite{DBLP:journals/ads/LiZZR24}. 

In this paper, we consider techniques that allow one to \textit{control the content of the augmentation}.
We experiment with Diffusion Models~\cite{DBLP:conf/nips/HoJA20}, a class of (Vision) Generative AI techniques that achieved state-of-the-art performance in image generation tasks~\cite{dm-survey-1}. 
These models operate by reversing a gradual noising process, starting from a simple distribution and iteratively refining it to generate high-quality images. 
For example, given an initial noisy image, the model denoises it step by step to produce a coherent image. Different randomly sampled noise seeds lead to variations in the generated images, allowing these models to create diverse and unique images. 

Conditional Diffusion Models allow for further control over image generation using different types of input conditioning. 
For example, Stable Diffusion~\cite{DBLP:conf/cvpr/RombachBLEO22} and DALL-E~\cite{DBLP:conf/icml/RameshPGGVRCS21} use 
\textit{single conditioning} through a textual description to guide the denoising process, as a form of requirement specification (e.g., ``\textit{generate a sunny driving image scenario.}''). 
This guidance concept aims to align the generated images closely with the conditions or descriptions provided. 
Other techniques use \textit{multiple conditioning}. For example, InstructPix2Pix~\cite{DBLP:conf/cvpr/BrooksHE23} takes as input an image and a textual editing instruction that describes the modification to be applied to the image, whereas ControlNet~\cite{DBLP:conf/iccv/ZhangRA23} facilitates the addition of arbitrary conditioning inputs to a pre-trained Stable Diffusion model.

\section{Solution}\label{sec:solution}

Our methodology seamlessly integrates into the standard system-level testing loop without requiring major modifications to the simulator or the ADS. 
The key idea lies in manipulating the environment perceived by the ADS through diffusion-based augmentation, while the actual driving commands are executed on the original simulator. 
Our methodology consists of three main phases, namely Domain Augmentation, Semantic Validation, and Knowledge Distillation.

The first phase (Domain Augmentation) involves intercepting the images captured by the car cameras. 
These images are then processed using diffusion models to generate a new image depicting the same road structure but a different ODD condition (such as background and weather conditions). 
The main reason is that a well-trained lane-keeping ADS should focus on the foreground features that characterize the road scenario, instead of the features in the background~\cite{nvidia-dave2}.

The second phase (Semantic Validation) involves a validation step of the generated image to assess the validity and semantic and label preservation between the original and the augmented image. 
Particularly, our approach aims to generate road images devoid of visual artifacts, distortions, inconsistencies, or hallucinations and that are \textit{semantically equivalent} to the original one in terms of geometrical features (e.g., direction, lanes, length, width). The semantic validation process is fundamental to maintaining the validity of the augmentation process, as significant changes to the road structure introduced during augmentation could cause the ADS to make decisions based on misleading information.
Our approach addresses visual and semantic consistency, whereas physics-related factors such as changes in friction and traction (e.g., due to snowy conditions) are not considered.

The third and last phase (Knowledge Distillation) enables the rendering of new ODDs online, during the execution of a simulation.
The diffusion models used in the first phase for domain augmentation are not suitable for online usage because they are too slow at inference time. Hence, we train a faster cycle-consistent generative neural network~\cite{DBLP:conf/iccv/ZhuPIE17}, using the output images produced by the first phase as the training set. At each simulation step, this network transforms the input image to reproduce the domain augmentation of the diffusion models. 
If the augmented image passes the semantic validation, it is forwarded to the ADS for processing.
The ADS processes the image and predicts the appropriate driving commands based on the augmented ODD. The predicted driving commands are sent to the simulator, which actuates them, completing the feedback loop. The simulator then modifies the virtual environment and provides the vehicle with updated sensor data, thus preparing for the next iteration of the testing loop.
In the next sections, we describe each step of each phase.

\subsection{Domain Augmentation}\label{sec:solution:genai}

We analyze three alternative controllable augmentation strategies based on categories of diffusion models to introduce environmental ODD changes in driving images: \textit{\instruct}, \textit{\inpaint}, and \textit{\refining}.

\begin{figure}[t]
\centering
\includegraphics[width=1\linewidth]{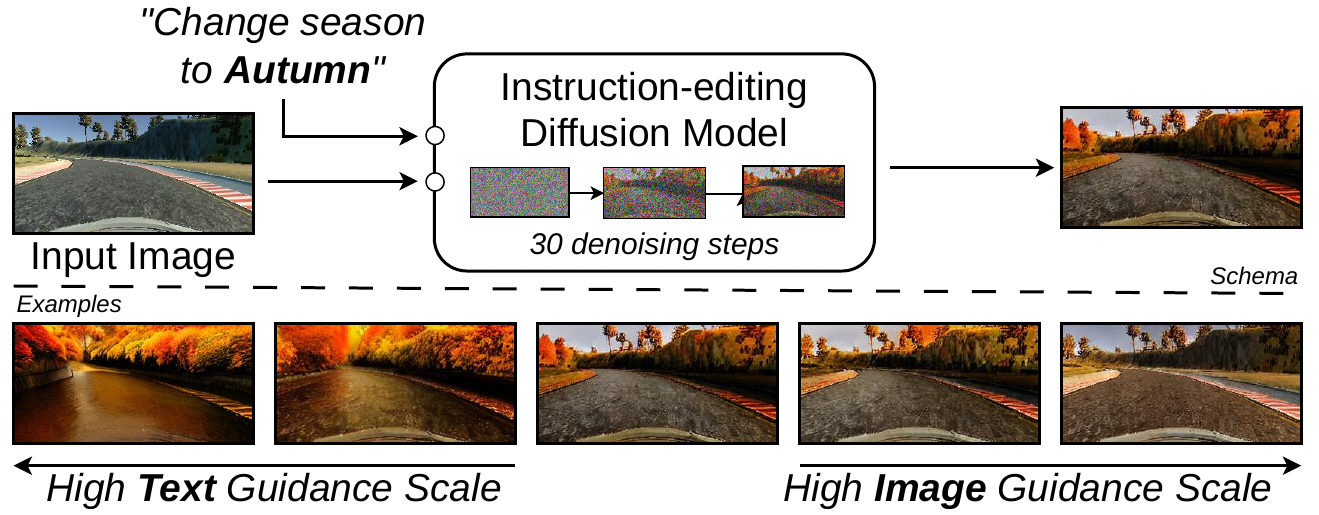}
\caption{Instruction-edited Domain Augmentation Strategy.}
\label{fig:solution:genai:instruct}
\end{figure}

\head{\instruct}
This category takes two inputs: the image to be modified and an editing instruction (e.g., ``add trees'' or ``change season to autumn''), and produces an output image with the editing instruction applied. 
Instances of \instruct models are InstructPix2Pix~\cite{DBLP:conf/cvpr/BrooksHE23} and  SDEdit~\cite{DBLP:conf/iclr/MengHSSWZE22}.
\instruct models can be configured with two parameters, \textit{image} and \textit{text guidance scale}, that represent how much the two inputs influence output generation. 
The first parameter \textit{image} dictates how much of the structure and spatial details of the input image should be preserved. The \textit{text guidance scale} determines the strength to use when applying the editing instruction. 
\autoref{fig:solution:genai:instruct} reports a schema of the strategy (top) and how the two guidance scales can influence the augmentation process (bottom).
Specifically, we can observe that an excessively high text guidance scale can compromise the road semantics, while overly increasing the image guidance scale too much may result in the insufficient application of the desired edit. 

\head{\inpaint}
This category employs a text-to-image diffusion model that performs inpainting. Instances of \inpaint models are Stable Diffusion~\cite{DBLP:conf/cvpr/RombachBLEO22}, DALL-E~\cite{DBLP:conf/icml/RameshPGGVRCS21}, and Pixart-$\alpha$~\cite{DBLP:conf/iclr/ChenYGYXWKLLL24}.

We customized the inpainting pipeline to preserve the parts of the images related to driving actions (e.g., the road), while the rest of the image can be ``repainted'' by the diffusion model. 
We identify the road automatically using a semantic mask that describes which pixels in the image belong to the semantic class \textit{road}. As the semantic mask is provided by the simulator, it ensures perfect semantic segmentation.

\begin{figure*}[t]
\centering
\includegraphics[width=1\linewidth]{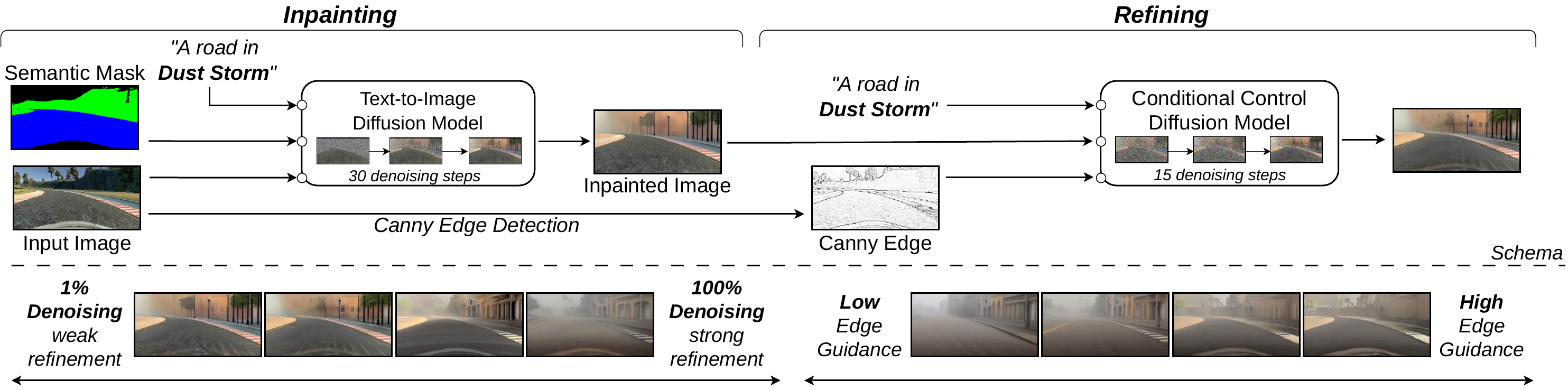}
\caption{\inpaint and \refining Domain Augmentation Strategies.}
\label{fig:solution:genai:refining}
\end{figure*}

The inpainting text-to-image model takes three inputs: an input image, a mask, and a textual prompt that describes the desired image.
The model generates an image by preserving only the content selected by the mask while guiding the entire image to align as closely as possible with the textual prompt. In our setting, this process ensures that only the parts outside the road are replaced with new content, thus maintaining the shape of the road since it is not modified by the inpainting strategy.
Note that the road in the inpainted image is the same as the one in the input image. For instance, in \autoref{fig:solution:genai:refining} (top left), the surrounding environment has been transformed, while the road structure, markings, and position are unchanged.

\head{\refining}
This model category adds a step to the \textit{\inpaint} strategy by performing a refinement of the entire image. 
The goal is to improve the visual coherence between the preserved and the generated parts of the inpainted image.
While the traditional denoising process of diffusion models starts from fully-noised images, the refinement step starts from a partially-noised (inpainted) image.
This makes the refined image more similar to the initial inpainted one. The noise level removed during denoising determines the difference between the inpainted and final image; higher noise removal leads to greater differences.

The refinement step employs a different type of diffusion model compared to the inpainting step. This is because image generation with a text-to-image model (used during inpainting) can be guided using only text. This is sufficient for the inpainting step as the road semantics are ensured by the semantic mask that preserves the road. However, during the refinement step, the entire image is modified, and thus the shape of the road might change.
For this reason, in our study, we evaluated a category of diffusion model with conditional controls that allow to better guide the augmentation process, such as ControlNet~\cite{DBLP:conf/iccv/ZhangRA23} with Canny edge~\cite{DBLP:journals/pami/Canny86a} conditioning, or T2I Adapter~\cite{DBLP:conf/aaai/MouWXW0QS24}.
This model takes three inputs: the edge map derived from the original input image (captured in the simulator), a partially noisy inpainted image, and a textual prompt that describes the desired image. The model takes the noisy image and refines it using both the edge map and the textual prompt. The edge map ensures that the refined image retains edge structures similar to the original, while the textual prompt directs the overall content. 

\autoref{fig:solution:genai:refining} provides an overview of the \refining process (top), and reports some augmentations obtained with different levels of denoising and guidance scales for the same input image (bottom). A stronger refinement process (more denoising) results in significant differences from the initial image, but can also lead to inferior preservation of road semantics.
Similarly, higher edge guidance scale values preserve road semantics more, while lower values encourage greater freedom, diversity, and realism in the generated image.

\subsection{Semantic Validation}
\label{sec:solution:semantic}

Diffusion model categories aim to produce visually appealing outputs, but they may still generate invalid outputs, failing to preserve the semantics of the road during enhancement, for example, by widening the road or introducing new intersections. To mitigate this, our methodology includes a semantic validation step to minimize incorrect augmentations.

The main objective of this phase is to check that the generated augmentation is characterized by a road that is semantically equivalent to the road in the image captured in the simulator. 
To this end, we filter out augmentations that do not preserve the original road semantics using the OC-TSS metric (One Class - Targeted Semantic Segmentation)~\cite{2024-Lambertenghi-ICST}.

OC-TSS is a similarity metric that measures semantic details and structural differences between two images by focusing on a single task-relevant class within the semantic masks predicted by a fine-tuned segmentation model.
This metric ranges from 0 to 1, where 1 indicates perfect semantic equivalence between the original and augmented images, and 0 suggests complete dissimilarity.
OC-TSS has been applied in previous work~\cite{2024-Lambertenghi-ICST} to assess the accuracy of Generative AI models in translating images across domains for a lane-keeping ADS. In line with this study, our analysis focuses on the semantic class ``road'', as road lanes represent the relevant image characteristics for a lane-keeping ADS. Unlike the original work, we employ a U-Net architecture~\cite{DBLP:conf/miccai/RonnebergerFB15} for semantic segmentation, rather than the SegFormer architecture~\cite{SegFormer}, for its computational efficiency.

\begin{figure}[t]
    \centering
    \includegraphics[width=\linewidth]{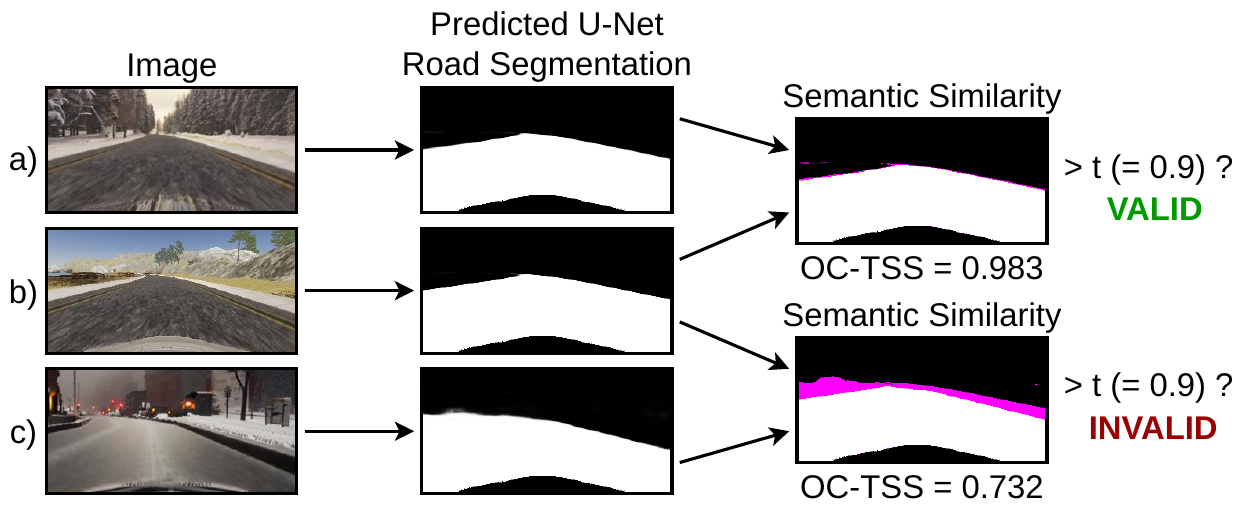}
    \caption{Semantic Validation using OC-TSS~\cite{2024-Lambertenghi-ICST}.}
    \label{fig:solution:semantic}
\end{figure}

\autoref{fig:solution:semantic} illustrates our semantic validation process.
The top row of \autoref{fig:solution:semantic}~(a) shows a semantically valid augmentation of the input image (\autoref{fig:solution:semantic}~(b), while the bottom row reports an incorrect augmentation (\autoref{fig:solution:semantic}~(c). Specifically, the augmentation in the bottom figure is considered semantically invalid because the road's orientation changes to the left instead of continuing straight as in the original image.
The middle column presents the semantic segmentation masks computed by the U-Net model for each image. In these masks, the road is represented in white, and the background is depicted in black.
The final step of our process, represented in the right column, involves measuring the distance between the semantic masks of the augmented and original images, the differences highlighted in pink. Augmentations with a similarity score below a threshold are discarded. 
Higher threshold values ensure validity, but they may discard valid images and increase the time required to generate a semantically valid augmentation. However, lower thresholds could compromise testing by allowing too many invalid images. 

In this study, we determine the threshold based on empirical observations to balance filtering out invalid augmentations and retaining enough variability for thorough ADS testing (\autoref{sec:evaluation:setup}).

\subsection{Knowledge Distillation}
\label{sec:solution:neural}
The final phase involves creating a fast and consistent neural rendering engine in the simulator, using outputs from the previous phases. The diffusion-based models, while effective for diversity, are computationally expensive and can produce potentially inconsistent augmentations, which may be problematic for the temporal coherence of a simulation.

Therefore, we adopt a technique known as knowledge distillation~\cite{DBLP:journals/corr/HintonVD15}, where a smaller model (\textit{student}) is trained to replicate the behavior of a larger, more complex model (\textit{teacher}, i.e., the diffusion model). 
In particular, for the student model, we use a cycle-consistent generative network~\cite{DBLP:conf/iccv/ZhuPIE17} that can map images from the original domain (virtual images from the simulator) to another domain (augmented images by the diffusion models). Instances of these architectures are CycleGAN~\cite{DBLP:conf/iccv/ZhuPIE17} or UNIT~\cite{DBLP:conf/nips/LiuBK17}.
This technique is widely used for image-to-image translation tasks, including the autonomous driving domain, due to its low computational overhead, which makes it suitable for runtime usage in simulators~\cite{2022-Stocco-TSE,deeproad}. Moreover, training a separate student model for each domain allows effective learning of the key aspects of the teacher model's output, enhancing rendering consistency.

This strategy involves first training the cycle-consistent network to learn the mapping between the original and augmented image domains produced by the diffusion models. This process, while computationally intensive, is performed only once for each domain. Then, during the online system-level testing of the ADS, the trained cycle-consistent network model generator translates images at runtime, i.e., during the execution of the simulation.

An important advantage of this approach is that it does not require collecting new data to train the model, as it leverages existing pre-trained diffusion models. This means that the process can be easily automated and does not require human intervention (e.g., collecting and labeling data), making it an efficient solution for rapidly generating consistent and high-quality domain augmentations online, during ADS simulation.

\section{Evaluation}\label{sec:evaluation}


We evaluated the proposed approach through the following research questions:

\noindent
\textbf{RQ\textsubscript{1} (semantic validity and realism):}
\textit{Do diffusion models generate images that are semantically valid and realistic? How is the semantic validator at detecting invalid augmentations?}

\noindent
\textbf{RQ\textsubscript{2} (effectiveness):}
\textit{How effective are augmented images in exposing faulty system-level misbehaviors of ADS?}

\noindent
\textbf{RQ\textsubscript{3} (efficiency):}
\textit{What is the overhead introduced by diffusion model techniques in simulation-based testing? Does the knowledge-distilled model speed up computation?}

\noindent\textbf{RQ\textsubscript{4} (generalizability):}
\textit{Does the approach generalize to complex urban scenarios and multi-modal ADS?}

The first research question aims to assess the semantic validity of the augmentations generated by our methodology. Specifically, the focus is on whether diffusion models can transform images while preserving road semantics, and whether the proposed semantic validator can effectively identify roads with different semantics.
The second research question aims to check the utility of the proposed approach in identifying potential faults in ADS that may not be detected using only the simulator's capabilities.
The third research question evaluates the computational cost of our approach, which is crucial to understanding scalability in real-world ADS testing scenarios.
The fourth research question studies the generalizability of our approach to test multi-modal ADS in complex urban environments with pedestrians, vehicles, and traffic lights.

\subsection{Experimental Setup}\label{sec:evaluation:setup}

In this section we describe the experimental setup used for RQ\textsubscript{1}, RQ\textsubscript{2}, RQ\textsubscript{3}, while the setup changes required for RQ\textsubscript{4} are described  in Section~\ref{sec:evaluation:RQ4}, before presenting the results.

\head{Simulation Platform}
We carried out our evaluation using the Udacity simulator with behavioral cloning ADS models~\cite{udacity-simulator}, a widely adopted platform in the literature~\cite{DBLP:conf/icse/0001WCT20,udacity-challenge}. 
The simulator supports various closed-loop tracks (divided into $40$ sectors) to test behavioral cloning ADS, including a predefined set of ODDs, such as different times of day/night and three weather conditions (rainy, snowy, and foggy)~\cite{2020-Stocco-ICSE}.
We extended the simulator to generate semantic segmentation masks for vehicle camera images of size 160$\times$320 (height $\times$ width) to identify the regions for inpainting.
In addition, we developed a synchronous simulation mechanism that pauses the simulation during image augmentation and resumes once the new image is generated. This ensures that the augmentation process is transparent to the system-level testing process.

\head{Lane-keeping ADS}
We evaluated four different single-camera, lane-keeping DNN-based ADS as systems under test to assess the performance of the proposed methodology.
In particular, we selected Nvidia \davetwo~\cite{DBLP:journals/corr/BojarskiTDFFGJM16}, Chauffeur~\cite{chauffeur} and Epoch~\cite{epoch} since they have been often used in multiple testing works~\cite{DBLP:conf/icse/TianPJR18,2020-Stocco-ICSE,DeepGuard,deepxplore}. Finally, we also included a recent architecture based on Vision Transformer (ViT-based)~\cite{vit-based-nn} that achieved state-of-the-art performance in lane-keeping tasks.

\begin{table}[]
\caption{ODD Domains.}
\label{tab:evaluation:setup:domains}
\centering
\begin{tabularx}{.48\textwidth}{@{}p{.055\textwidth}X@{}}
\toprule
\textbf{Category} & \textbf{Domains} \\ \midrule
Weathers          & cloudy, dust storm, foggy, lightnings, overcast, smoke, sunny \\ \midrule
Seasons           & autumn, spring, summer, winter \\ \midrule
Daytimes         & afternoon, dawn, dusk, evening, morning, night, sunset \\ \midrule
Locations         & coast, desert, forest, lake, mountain, plains, rivers, rural, seaside \\ 
\midrule
Cities            & beijing, berlin, chicago, el cairo, london, new york, paris, rome, san francisco, sidney, tokyo, toronto \\ \midrule
Countries         & australia, brazil, canada, china, england, france, germany, italy, japan, mexico, morocco, usa \\ \bottomrule
\end{tabularx}
\end{table}


\head{Operational Design Domains Selection}
We selected ODDs that encompass diverse conditions from existing standards (see \autoref{sec:background:testing}). 
We filtered out domains that do not preserve the driving action when applied for domain augmentation. For example, when converting a sunny road image to a snowy condition, it might require the prediction of a different steering angle that accounts for the different friction, despite the road being the same. 
Overall, we identified 6 domain categories and 52 distinct label-preserving domains (\autoref{tab:evaluation:setup:domains}).

To provide further insight into the difficulty of these domains, we measured the challenge they pose by computing the distance between augmented domains and the training data. This was done using the reconstruction error of a Variational Autoencoder (VAE)~\cite{DBLP:journals/corr/KingmaW13} to categorize the domains into three clusters: in-distribution domains (closer to the training distribution, e.g., familiar domain or road conditions), in-between domains (moderately different from the training distribution), and out-of-distribution domains (significantly different from the training distribution). In-distribution domains are useful for testing the robustness of the ADS by simulating scenarios similar to those that the model has previously encountered. On the other hand, out-of-distribution domains challenge the ADS's ability to generalize to new, unfamiliar conditions that are not present or rarely available in training data.

To determine the classification of these domains, we first trained the VAE to reconstruct the training data. The lower the reconstruction error, the closer the domain is to the training distribution. We generated $2,000$ augmented images for each domain using the three domain augmentation strategies and measured the reconstruction error for each. The domains were then sorted by reconstruction error and categorized as in-distribution, in-between, or out-of-distribution.
Finally, we selected three domains that were categorized in the same group for all three augmentation techniques. The final selections were as follows: for in-distribution domains, we included sunny, summer, and afternoon conditions; for in-between domains, we chose autumn, desert, and winter; and for out-of-distribution domains, we selected dust storm, forest, and night scenarios.

\head{Diffusion Models Calibration}
We used three state-of-the-art pre-trained diffusion models for our categories: InstructPix2Pix~\cite{DBLP:conf/cvpr/BrooksHE23} for \instruct, Stable Diffusion~\cite{DBLP:conf/cvpr/RombachBLEO22} for \inpaint, and ControlNet~\cite{DBLP:conf/iccv/ZhangRA23} with Canny edge~\cite{DBLP:journals/pami/Canny86a} conditioning for \refining.
We fine-tuned the hyperparameters of each considered model before answering the research questions. We prioritized image fidelity, adherence to instructions, and preservation of essential road features, which will be evaluated in our first research question.
In particular, we configured all diffusion models to use the UNIPC multistep scheduler~\cite{DBLP:conf/nips/ZhaoBR0L23} with 30 inference denoising steps as \textit{noise sampling strategy}.
We chose UNIPC because it focuses on generating good images with a few denoising steps. In our exploratory experiments, a higher number of steps did not lead to significantly better images, but only introduced additional computational overhead.

For InstructPix2Pix, we set the image guidance scale to~$2$ and the text guidance scale to $10$. These settings were found to be a good balance between image and instruction inputs, preserving key features from both sources. 
In the Stable Diffusion inpainting pipeline, we used a text guidance scale of $10$, which maintained a high level of control over the generated content without compromising the quality of the inpainting process.
ControlNet refining was configured with a text guidance scale of $10$ and a noise level of $50\%$. This configuration preserved the structural integrity of the road while still allowing for meaningful and diverse augmentations. Higher noise levels were found to risk excessive alteration of images and the potential loss of essential road semantics.

\head{Semantic Validator Configuration}
To determine an appropriate threshold for the OC-TSS metric, we collected $150$ images from the simulator with different semantics, which were manually assigned to three categories: images of straight roads, right turns, and left turns. We computed the OC-TSS for all images and evaluated the similarity between images within the same category and across different categories.

We aimed for a threshold that considers images within the same category as semantically similar and those in different categories as distinct, giving higher priority to filtering out invalid images that belong to another category, rather than including as many valid images from the same category as possible. Consequently, after manual inspection of a sample of included/excluded images, we chose a conservative threshold of $0.9$, which minimizes the inclusion of semantically incorrect augmentations, while at the same time avoiding the exclusion of too many valid images. Thus, augmentations with an OC-TSS $\geq 0.9$ are considered semantically consistent with the original layout of the road by our automated semantic validator, while those below 0.9 are considered invalid and discarded.

\head{Knowledge Distillation Configuration}
We used CycleGAN as the student model to distill knowledge from the pre-trained diffusion models. The CycleGAN architecture followed the recommendations in \cite{DBLP:conf/cvpr/KarrasLAHLA20} to reduce droplet artifacts.

We trained one CycleGAN for each of the nine selected ODDs and the three augmentation strategies, resulting in 27 models. Each model was trained for 10 epochs using $2,000$ pairs of images (from the simulator and augmented), with checkpoints saved at the end of every epoch. The best checkpoint was selected according to the Fréchet Inception Distance~\cite{DBLP:conf/nips/HeuselRUNH17}, a metric that measures the distance between two sets of images (the ones generated by CycleGAN and the ones generated by the diffusion models) by comparing their feature distributions.

\head{Hardware and Software}
All experiments were executed on a server with an AMD 5950X CPU, 64 GB RAM, and two Nvidia 4090 GPUs (24 GB VRAM each). The software environment includes Python 3.10, CUDA 12.1 for GPU acceleration, Pytorch 2.3.0 for ADS implementations, and Huggingface \texttt{diffusers} 0.27.2 for the diffusion models.

Our evaluation required more than $1,000$ GPU hours and involved 2.5 million image pairs generated over $52$ ODD domains using $3$ augmentation techniques. This process included filtering the domains to keep the experiment manageable within a reasonable time frame, as training $36$ CycleGAN models for $10$ epochs each took more than $150$ GPU hours.

\subsection{Metrics}\label{sec:evaluation:metrics}

\head{Semantic Validator Effectiveness}
We consider valid augmentations as the positive class and invalid augmentations as the negative class. Correspondingly, a True (resp. False) Positive TP (resp. FP) is an image regarded as a valid augmentation by our semantic validator, which is valid (resp. invalid) according to the ground truth. Similarly, a True (resp. False) Negative TN (resp. FN) is an image regarded as an invalid augmentation by our semantic validator, which is invalid (resp. valid) according to the ground truth.
To assess the effectiveness of our semantic validation methodology, we utilize the confusion matrix
[[TP, FP], [FN, TN]], either with absolute or percentage values.

\head{Testing Effectiveness}
We utilize two categories of metrics to evaluate ADS performance at the system level: one for measuring misbehavior and another for assessing driving quality. The first category directly quantifies errors, including incidents in which the vehicle deviates from the lane boundaries (Out-of-Bounds, OOB) or collides with obstacles (C). We use Failure Track Coverage (FTC) to capture the spatial distribution of errors, which identifies the percentage of track sectors where misbehaviors occur. This metric indicates whether errors are concentrated in specific challenging areas or spread over the entire track. In this way, we can determine whether errors arise primarily from the complexity of specific track sections or are induced more broadly by the new domain.

We assess two key metrics for driving quality relative to the nominal behavior of the ADS. The first is the Relative Cross-Track Error (RCTE), which measures the ratio of the average distance from the lane center in the test domain compared to the nominal domain. An RCTE value greater than 1 indicates degraded performance (the vehicle is closer to the edge of the road), while a value less than 1 indicates improved position accuracy (the vehicle is closer to the center of the road). The second metric is Relative Steering Jerk (RSJ), which calculates the difference in the rate of change of steering angle between the test and nominal domains. Higher RSJ values suggest more abrupt steering adjustments, while lower values indicate smoother driving. Although these metrics do not directly indicate errors, they provide valuable insight into potential performance degradation caused by specific domains.

\head{Computational Overhead}
We evaluate the computational overhead of our domain augmentation strategies by measuring the average time required to generate the augmented image and the average time required to complete our experiments, including both the augmentation process and the subsequent testing of the ADS. We compare these timings with a baseline where no augmentation is applied, allowing us to quantify the additional computational load introduced by each strategy. 

\subsection{RQ\textsubscript{1}: Semantic Validity and Realism}
\label{sec:evaluation:RQ1}

\begin{table}[t]

\caption{RQ\textsubscript{1}: Semantic Validity Confusion Matrix.}
\label{fig:evaluation:rq1:semantic_validity}

\renewcommand{\arraystretch}{0.9}
\setlength{\tabcolsep}{10pt}

\centering

\begin{tabular}{@{}ccc@{}}

\toprule

& \textbf{Valid Augmentation} 
& \textbf{Invalid Augmentation} \\ 

\midrule




\begin{tabular}[c]{@{}c@{}}Predicted \\ Valid Augmentation\end{tabular}   & \begin{tabular}[c]{@{}c@{}}48\\ (55\%)\end{tabular} & \begin{tabular}[c]{@{}c@{}}3\\ (3\%)\end{tabular}   \\ 

\midrule

\begin{tabular}[c]{@{}c@{}}Predicted \\ Invalid Augmentation\end{tabular} & \begin{tabular}[c]{@{}c@{}}16\\ (18\%)\end{tabular} & \begin{tabular}[c]{@{}c@{}}20\\ (23\%)\end{tabular} \\ 

\bottomrule

\end{tabular}
\end{table}

We conducted two surveys with human assessors to evaluate the semantic validity of the images generated by the diffusion models, as well as their degree of realism. We recruited participants from Amazon Mechanical Turk (MTurk) and personal contacts using convenience sampling~\cite{conveniencesampling}.
Each MTurk participant answered $200$ questions, while the others answered $100$ questions. In the two studies, we collected $5,300$ responses, of which only $4,500$ were retained due to failure to answer the control questions of our surveys. Ultimately, we retained responses from $35$ participants, of which $10$ from MTurk and $25$ from personal contacts. 

\subsubsection{Semantic Validity}
Participants were shown two randomly ordered images and asked whether the images represented the same semantics of the road, focusing on the shape of the road and the direction of turn. 
For this study, we randomly selected $36$ pairs of images for each of the three domain augmentation strategies (\instruct, \inpaint, and \refining). This resulted in a total of $108$ pairs, with each strategy contributing $18$ semantically valid transformations and $18$ invalid transformations, according to our semantic validator. In addition, we included two control questions to filter out low-quality responses: one where the road was the same and one where the road was entirely different. In total, the first questionnaire contained $110$ questions, of which two were used for quality checks.

We considered a road semantically equal (or different) when at least $\frac{2}{3}$ of the participants agreed on the outcome. Overall, the participants reached a consensus on $80.6\%$ of the pairs (87 out of 108).
Our study revealed a positive correlation between OC-TSS similarity scores and user opinions, with a Pearson correlation coefficient of $0.63$ ($p$-value=$2.63\cdot10^{-13}$).

\autoref{fig:evaluation:rq1:semantic_validity} presents the results as a confusion matrix, where columns represent the judgments of human participants and rows show the results of the semantic validator, with valid augmentations considered the positive class.
Our semantic validator failed to filter out invalid generations in only $3\%$ of the cases (3 FPs), while rejecting $18\%$ of the valid transformations (16 FNs).
The former error can affect the effectiveness of the proposed methodology as it could lead to testing ADS on images with different semantics, potentially compromising the validity of ADS testing. The latter, while less severe, may increase the time required to find valid augmentations.

\subsubsection{GenAI Realism}

In the second study, we evaluated the realism of the augmented images. Participants were presented with individual images and asked to rate their realism on a 5-point scale ranging from $1$ (not realistic) to $5$ (very realistic).
We selected $18$ semantically valid transformations for each of the three domain augmentation strategies. We also included $18$ images from the simulator and $18$ real-world driving images for a better comparison. In total, the second questionnaire contained $90$ questions. We computed the average realism score for each category of images (augmented, simulator, and real-world) based on the participants' ratings.

\begin{figure}[t]
    \centering
    \includegraphics[width=.95\linewidth]{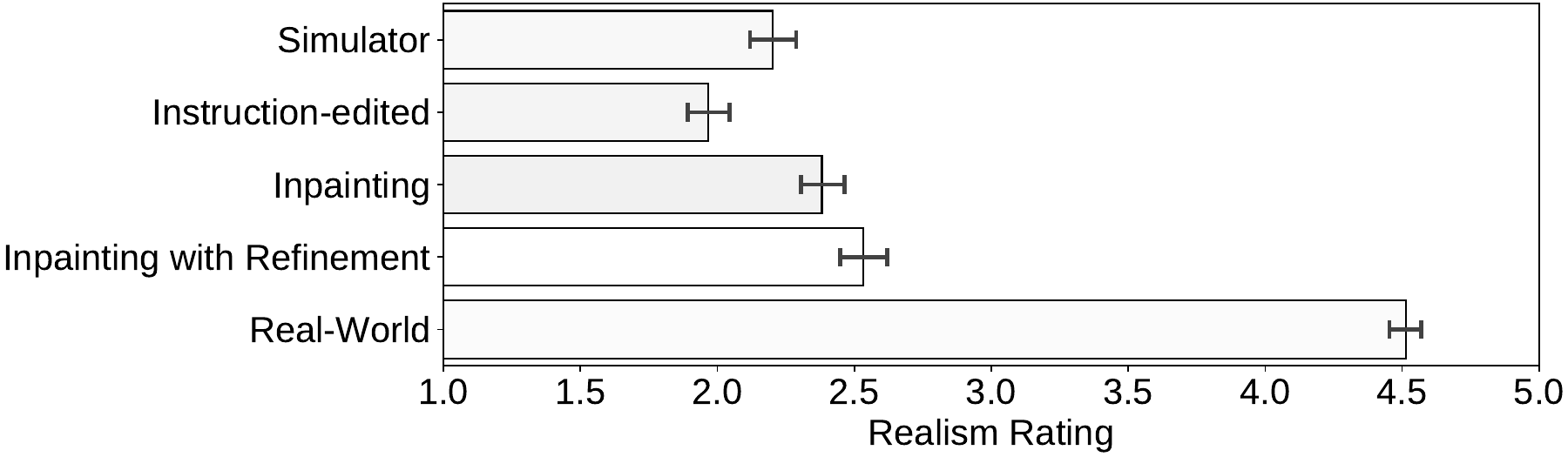}
    \caption{RQ\textsubscript{1}: Realism.}
    \label{fig:evaluation:realism}
\end{figure}

\autoref{fig:evaluation:realism} shows the results. The images generated by \textit{\inpaint} and \textit{\refining} strategies are perceived as more realistic compared to those generated by the \textit{\instruct} strategy. The \textit{Mann-Whitney U test} with an alpha of $0.05$ indicates statistically significant differences in realism scores between strategies, with a medium effect size ($0.5$ and $0.6$, respectively). Furthermore, \textit{\refining} was found to produce more realistic images than \textit{\inpaint} ($p$-value=$0.02$), with a small effect size ($0.17$).

\begin{tcolorbox}[boxrule=0pt,frame hidden,sharp corners,enhanced,borderline north={1pt}{0pt}{black},borderline south={1pt}{0pt}{black},boxsep=2pt,left=2pt,right=2pt,top=2.5pt,bottom=2pt]
\textbf{RQ\textsubscript{1}}: \textit{
The output of our automated semantic validator matched human judgment, with only $3\%$ of the augmented images incorrectly regarded as valid by the validator (and $18\%$ valid augmentations incorrectly discarded by the validator). Inpainting with Refinement is the augmentation approach that produces the most realistic images.}
\end{tcolorbox}

\begin{table*}[t]
\caption{RQ\textsubscript{2}: Effectiveness results for system-level testing on augmented domains.}
\label{fig:evaluation:rq2:aug_domains}
\footnotesize
\setlength\tabcolsep{6pt} 
\renewcommand{\arraystretch}{0.9}
\centering
\begin{tabular}{@{}lccccccccccccccc@{}}
\toprule

& \multicolumn{5}{c}{\textbf{In-distribution domains}} & \multicolumn{5}{c}{\textbf{In-between domains}} & \multicolumn{5}{c}{\textbf{Out-of-distribution domains}} \\ 

\cmidrule(r){2-6}
\cmidrule(r){7-11}
\cmidrule(r){12-16}
                            & C        & OOB         & FTC          & RCTE       & RSJ        & C        & OOB        & FTC         & RCTE      & RSJ       & C        & OOB         & FTC           & RCTE        & RSJ         \\ \midrule
\textbf{\davetwo}  &            &           &             &            &            &           &          &            &           &           &            &           &              &             &             \\
\quad \instruct                 & 0          & 4         & 5.0\%        & 1.13       & 0.85       & 10        & 19       & 45.0\%      & 1.62      & 0.61      & 8          & 66        & 82.5\%        & 1.90        & 0.70        \\
\quad\inpaint                  & 0          & 2         & 2.5\%        & 1.10       & 1.06       & 0         & 0        & 0.0\%       & 1.06      & 0.95      & 3          & 7         & 12.5\%        & 1.21        & 1.06        \\
\quad\refining                 & 0          & 0         & 0.0\%        & 0.75       & 2.37       & 0         & 0        & 0.0\%       & 0.90      & 0.68      & 0          & 4         & 5.0\%         & 0.96        & 0.80        \\ [0.5em] 
\textbf{Chauffeur} &            &           &             &            &            &           &          &            &           &           &            &           &              &             &             \\
\quad\instruct                 & 0          & 7         & 12.5\%       & 1.32       & 0.68       & 4         & 19       & 35.0\%      & 1.60      & 0.63      & 7          & 67        & 70.0\%        & 1.60        & 0.57        \\
\quad\inpaint                  & 0          & 0         & 0.0\%        & 0.93       & 0.82       & 1         & 2        & 7.5\%       & 1.23      & 0.80      & 1          & 7         & 12.5\%        & 1.05        & 0.80        \\
\quad\refining                 & 1          & 3         & 7.5\%        & 1.03       & 0.71       & 0         & 0        & 0.0\%       & 0.93      & 0.68      & 4          & 6         & 20.0\%        & 1.39        & 0.80        \\ [0.5em] 
\textbf{Epoch}     &            &           &             &            &            &           &          &            &           &           &            &           &              &             &             \\
\quad\instruct                 & 3          & 11        & 25.0\%       & 2.24       & 0.67       & 3         & 54       & 75.0\%      & 2.28      & 0.52      & 10         & 70        & 87.5\%        & 2.18        & 0.41        \\
\quad\inpaint                  & 0          & 3         & 7.5\%        & 1.62       & 0.86       & 0         & 5        & 7.5\%       & 1.65      & 0.71      & 3          & 30        & 52.5\%        & 1.98        & 0.71        \\
\quad\refining                 & 0          & 8         & 12.5\%       & 1.55       & 0.67       & 4         & 2        & 12.5\%      & 1.71      & 0.67      & 4          & 10        & 15.0\%        & 1.80        & 0.62        \\ [0.5em] 
\textbf{ViT-based} &            &           &             &            &            &           &          &            &           &           &            &           &              &             &             \\
\quad\instruct                 & 2          & 8         & 12.5\%       & 1.15       & 1.30       & 2         & 13       & 27.5\%      & 1.23      & 2.45      & 3          & 21        & 45.0\%        & 1.77        & 3.40        \\
\quad\inpaint                  & 1          & 6         & 12.5\%       & 1.21       & 1.74       & 1         & 15       & 25.0\%      & 1.21      & 1.85      & 0          & 10        & 17.5\%        & 1.07        & 2.02        \\
\quad\refining                 & 2          & 8         & 12.5\%       & 1.05       & 1.79       & 2         & 12       & 25.0\%      & 1.17      & 1.77      & 2          & 20        & 22.5\%        & 1.02        & 1.02        \\ 
\bottomrule
\end{tabular}
\end{table*}

\subsection{RQ\textsubscript{2}: Effectiveness}\label{sec:evaluation:RQ2}

This experiment evaluates the effectiveness of the domain augmentation approaches in discovering errors in lane-keeping ADS.
The experiment aims to determine how many errors each approach could find and the nature of these errors.
For each strategy, we ran $2,000$ ADS simulation steps for each augmented domain, collecting the failures and driving quality metrics (\autoref{sec:evaluation:metrics}).
As a baseline, we used the predefined set of domains available within the simulator (\autoref{sec:evaluation:setup}).

\autoref{fig:evaluation:rq2:aug_domains} reports the results with the augmented domains, and \autoref{fig:evaluation:rq2:sim_domains} shows the domains available in the simulator.
The findings indicate that the proposed methodology can identify misbehaviors in all four ADS. 
As expected, domains similar to training conditions (in-distribution) showed fewer errors and lower failure track coverage than in-between or out-of-distribution domains, reflecting the increased difficulty due to domain shifts. 
However, even for \textit{in-distribution} domains, our approach revealed failures, from a total of $6$ OOB incidents (\davetwo) to $5$ collisions and $22$ OOB incidents (ViT-based). The maximum failure track coverage with in-distribution domains was $25\%$. 
Failure track coverage increased as the domains deviated further from the training set. For example, when Epoch was tested with in-between domains, new errors were seen in up to $75\%$ of the sectors, rising to $87.5\%$ with out-of-distribution domains.

Simulated domains revealed errors in up to $17.5\%$ of the track, while augmented domains reached $87.5\%$. In-distribution domains generated by the augmentation strategies did not show more misbehaviors than those in the simulator. In contrast, in-between and out-of-distribution domains triggered more errors, especially with the \instruct strategy. Although these scenarios occur less often in training, the generated roads remain semantically validated and sufficiently representative for valid testing.

Furthermore, we found that the \instruct strategy was the most effective across all three domain sets (in-distribution, in-between, and out-of-distribution). 
This effectiveness can likely be attributed to two main factors.
First, the domains generated using instruction-editing were perceived as less realistic in our human study (\autoref{sec:evaluation:RQ1}). 
Despite being semantically validated by our automated validator, these less realistic images may still mislead the ADS into making incorrect decisions. 
Second, the average distance from these domains to the training domains was higher compared to that generated by the other two domain augmentation strategies. 
Specifically, the average reconstruction errors for the \instruct strategy ranged from $0.074$ (in-distribution) to $0.218$ (out-of-distribution), while the errors for the \inpaint strategy ranged from $0.067$ to $0.078$, and for the \refining strategy, they ranged from $0.070$ to $0.082$.

\begin{table}[t]
\caption{RQ\textsubscript{2}: Effectiveness results for system-level testing on domains available in the Udacity simulator.}
\label{fig:evaluation:rq2:sim_domains}
\footnotesize
\setlength\tabcolsep{14pt} 
\renewcommand{\arraystretch}{0.9}
\centering
\begin{tabular}{@{}lccccc@{}}
\toprule
                   & \multicolumn{5}{c}{\textbf{Simulator Domains}} \\ 
                   \cmidrule(r){2-6}
          & C       & OOB       & FTC       & RCTE      & RSJ         \\ \midrule
\textbf{\davetwo}  & 1         & 6       & 12.5\%   & 1.19      & 0.96        \\
\textbf{Chauffeur} & 2         & 9       & 17.5\%   & 1.20      & 0.99      \\
\textbf{Epoch}     & 0         & 5       & 10.0\%   & 1.68      & 1.01      \\
\textbf{ViT-based} & 1         & 7       & 17.5\%   & 1.21      & 1.22      \\ \bottomrule
\end{tabular}
\end{table}

We observed different behaviors between ADS models based on Convolutional DNNs (\davetwo, Chauffeur, and Epoch) and the one based on Vision Transformer. For the first ADSs, our augmented domains generally led to lower RSJ, indicating smoother steering responses. In contrast, the ViT-based ADS showed an increase in RSJ as domain distance increased, suggesting more abrupt steering adjustments in response to unfamiliar scenarios. These differences can likely be attributed to how these two types of neural network architectures process data. Convolutional DNNs, with their hierarchical structure and localized receptive fields, tend to focus on local features, which may allow for more stable and incremental responses. In contrast, ViTs, which utilize global attention mechanisms, can capture broader contextual information but may also be more sensitive to domain shifts, leading to more pronounced reactions to unfamiliar data. 

To assess whether the augmented data can enhance the robustness of ADS, we retrained the ADS using a combination of the original training data and the augmented ones. We fine-tuned the driving models for $50$ additional epochs, using early stopping with a patience of 20 epochs based on improvement in validation loss.
Empirical results show that, across the simulator domains, retrained ADS models showed, on average, a $35.2\%$ misbehavior reduction, up to $81.7\%$ in specific scenarios (i.e., foggy weather conditions). 

\begin{tcolorbox}[boxrule=0pt,frame hidden,sharp corners,enhanced,borderline north={1pt}{0pt}{black},borderline south={1pt}{0pt}{black},boxsep=2pt,left=2pt,right=2pt,top=2.5pt,bottom=2pt]
\textbf{RQ\textsubscript{2}}: \textit{
The proposed augmentation technique has been able to expose failures of four existing ADS models, even in domains close to the training one. It represents a valuable complement to the execution of tests in domains supported by the simulator, as it was able to discover failures in sectors that the simulator deemed failure-free.
}   
\end{tcolorbox}

\subsection{RQ\textsubscript{3}: Efficiency}\label{sec:evaluation:RQ3}

In this experiment, we assessed the overhead introduced by domain augmentation strategies, particularly focusing on the impact of large diffusion models and the potential efficiency gains from a knowledge-distilled model. 

\autoref{fig:evaluation:rq3:overhead} presents the results of testing \davetwo, with similar results observed for other lane-keeping ADS systems. A test run of \davetwo, consisting of $2,000$ simulation steps without augmentation, took approximately $15.7$ minutes (about $471.0$ms per simulation step). \davetwo required only $1.2$ms per prediction, while the remaining time was consumed by infrastructure tasks such as communication between the simulator and the agent, image processing, persistent logging, and simulation management. Other ADS systems showed inference times ranging from $1.1$ to $2.0$ ms.

The use of diffusion models significantly increased the test duration. Specifically, the \inpaint strategy extended the testing time to $53.2$ minutes (more than three times longer than the baseline), while the \instruct and \refining strategies increased it beyond $70$ minutes (more than four times longer). Although the \instruct strategy had a faster per-augmentation time ($894.7$ms), its overall testing duration was longer because our semantic validator detected a higher percentage of semantically invalid images and needed to be regenerated. Specifically, the semantic validator filtered out about 48\% of images augmented by \instruct, less than $1\%$ generated by \inpaint, and $12\%$ generated with \refining.

In contrast, the knowledge-distilled model based on a CycleGAN architecture significantly reduced the augmentation overhead. It generated images in just $12.3$ ms on average, resulting in a total testing time of approximately $16$ min, less than a $2\%$ increase over the baseline without augmentation.

\begin{table}[t]

\caption{RQ\textsubscript{3}: Performance Overhead.}
\label{fig:evaluation:rq3:overhead}
\renewcommand{\arraystretch}{0.9}
\setlength\tabcolsep{6pt} 
\centering

\begin{tabular}{
                    @{}
                    l
                    S[
                        table-number-alignment = center,
                        separate-uncertainty = true,
                        table-figures-uncertainty = 1
                    ]
                    S[
                        table-number-alignment = center,
                        separate-uncertainty = true,
                        table-align-uncertainty = true,
                        table-figures-uncertainty = 1,
                        table-figures-decimal = 1
                    ]
 @{}
 }
\toprule
                          & \textbf{\begin{tabular}[c]{@{}c@{}}Augmentation Time\\ (ms) \end{tabular}} & \textbf{\begin{tabular}[c]{@{}c@{}}Testing Run Time\\ (min) \end{tabular}} \\ \midrule
Baseline (no augmentation) &                                                                          & 15.7\pm0.0                                                                \\
\instruct                  & 894.7\pm0.8                                                               & 76.9\pm5.1                                                                \\
\inpaint                   & 1245.2\pm25.7                                                             & 53.2\pm1.2                                                                \\
\refining                  & 2172.6\pm29.1                                                             & 74.1\pm1.7                                                                \\
\ours                      & 12.3\pm0.7                                                                & 16.0\pm0.1                                                                \\ \bottomrule
\end{tabular}
\end{table}

\begin{tcolorbox}[boxrule=0pt,frame hidden,sharp corners,enhanced,borderline north={1pt}{0pt}{black},borderline south={1pt}{0pt}{black},boxsep=2pt,left=2pt,right=2pt,top=2.5pt,bottom=2pt]
\textbf{RQ\textsubscript{3}}: \textit{
Knowledge distillation is an essential component of our approach to achieve high simulation efficiency. The augmentation overhead without knowledge distillation is 470\% for \refining, the technique that produces more valid and more realistic images, which becomes just 2\% with knowledge distillation.}
\end{tcolorbox}

\subsection{RQ\textsubscript{4}: Generalizability} \label{sec:evaluation:RQ4}

To assess how well our methodology can generalize to complex and realistic scenarios, we performed a preliminary evaluation using the CARLA simulator~\cite{carla}, a widely used high-fidelity platform for autonomous driving research~\cite{drivefuzz,10.1145/3597926.3598072,liETAL2020}. We focused on scenarios from the CARLA Leaderboard (sensors track)~\cite{DBLP:journals/corr/abs-2405-01394} that require the ADS to perform multiple tasks such as lane-keeping, overtaking, obeying traffic signals, stop signs, and detecting and avoiding pedestrians.

We used InterFuser~\cite{DBLP:conf/corl/ShaoW00022} as the system under test, along with the pre-trained model provided in the original paper. InterFuser is an end-to-end ADS for urban driving with one of the highest scores in the CARLA leaderboard~\cite{carla-leaderboard} and was widely used in prior work~\cite{jia2023think,jia2023driveadapter,jaeger2023hidden}.
Technically, InterFuser processes multi-modal sensory inputs: three RGB camera views (front, left, and right) and LiDAR point cloud data, and outputs driving commands such as throttle, brake, and steering.

To assess driving quality and failures, we used several metrics from the CARLA leaderboard. They measure both the ability of the ADS to complete tasks and its adherence to traffic regulations: Driving Score (DS) reflects the overall ADS performance that combines achievements and penalties. Route Completion (RC) measures the percentage of the route completed. Penalties include Collisions with Pedestrians (CP) or Vehicles (CV), Off-Road Infractions (ORI), Red Light Infractions (RLI), and Stop Sign Infractions (SSI). 

To handle the multi-camera setup, we applied our methodology to three camera views and ensured consistency across them by using the same distilled model configured to augment each image.
Although our methodology does not operate on LiDAR point clouds, preserving the semantic consistency of the visual representations ensures that the LiDAR data by the simulator remain consistent with the augmented camera views.

Urban driving tasks, such as overtaking and collision avoidance, require the consideration of four additional semantic classes: pedestrians, vehicles, traffic signs, and traffic lights. These additional classes were addressed during both augmentation and validation. During augmentation, we configured a strategy based on inpainting to also preserve those parts of the image.
During validation, we applied OC-TSS to each semantic class and considered an augmentation valid only if all semantic classes were preserved within the threshold ($0.9$).

\begin{table}[t]
\caption{RQ\textsubscript{4}: System-level testing results with CARLA.}
\label{fig:evaluation:rq4:trans_effectiveness}
\footnotesize
\setlength\tabcolsep{3pt} 
\renewcommand{\arraystretch}{0.9}
\centering
\begin{tabular}{@{}lcccccccc@{}}
\toprule
{\textbf{CARLA Metrics}} 
                                  & DS        & RC        & CP        & CV        & ORI        & RLI        & SSI        \\ 
                                  & (\%)        & (\%)        & (\#)        & (\#)        & (\#)        & (\#)        & (\#)        \\ 
                                  \midrule
\textbf{InterFuser}               &           &           &           &           &            &            &            \\
\quad Baseline (no augmentation)  & 84.26     & 84.26     & 0         & 0         & 0          & 0          & 0  \\
\quad Augmented Domains           & 69.70     & 77.56     & 0         & 3      & 0          & 25       & 0  \\ \bottomrule
\end{tabular}
\end{table}

We replicated the experiments on the effectiveness of the augmented domains for failure exposure (RQ\textsubscript{2}) and measured the associated overhead (RQ\textsubscript{3}), using the same augmented domains as in the experiments with Udacity. 
CARLA supports various closed-loop urban maps for testing ADS. We considered Town05, one of the default maps provided by CARLA, with its default environmental configuration (e.g., sunny weather). Within Town05, we considered ten different scenarios (details are provided in the replication package).

\autoref{fig:evaluation:rq4:trans_effectiveness} reports the average effectiveness results in all scenarios and in all domains. As those scenarios are already designed to challenge the ADS, we provide the results with and without augmentation.
The results show that our approach successfully exposed new misbehaviors of the ADS, even in complex urban scenarios.

Regarding Route Completion (RC), InterFuser with no augmentation did not consistently achieve 100\%. Completion rates varied, with some runs at $98.5\%$ and others at $13.2\%$. Manual inspection of the logs revealed that the ego-vehicle gets ``stuck'' due to traffic jams caused by other vehicles that block intersections.
With the baseline simulator (no augmentation), no infractions and collisions were detected. With our neural augmentations, we discovered 25 previously unknown red light infractions and 3 previously unknown collisions with vehicles.

\begin{table}[t]

\caption{RQ\textsubscript{4}: Overhead in CARLA simulator. 
}
\label{fig:evaluation:rq4:trans_overhead}
\renewcommand{\arraystretch}{0.9}
\setlength\tabcolsep{4pt} 
\centering

\begin{tabular}{
                    @{}
                    l
                    S[
                        table-number-alignment = center,
                        separate-uncertainty = true,
                        table-figures-uncertainty = 1
                    ]
                    S[
                        table-number-alignment = center,
                        separate-uncertainty = true,
                        table-align-uncertainty = true,
                        table-figures-uncertainty = 1,
                        table-figures-decimal = 1
                    ]
 @{}
 }
\toprule
                          & \textbf{\begin{tabular}[c]{@{}c@{}}Augmentation Time\\ (ms) \end{tabular}} & \textbf{\begin{tabular}[c]{@{}c@{}}Testing Run Time\\ (min) \end{tabular}} \\ \midrule
Baseline (no augmentation) &                                                                          & 16.6\pm0.1                                                                \\
\instruct                  & 3377.7\pm28.9                                                               & 150.7\pm9.2                                                                \\
\inpaint                   & 8442.0\pm58.5                                                             & 358.5\pm16.2                                                                \\
\refining                  & 11350.5\pm137.8                                                             & 474.5\pm15.9                                                                \\
\ours                      & 82.0\pm5.7                                                               & 19.9\pm0.2                                                                \\ \bottomrule
\end{tabular}
\end{table}

Concerning the overhead (see Table~\ref{fig:evaluation:rq4:trans_overhead}), our methodology provides significant benefits also in CARLA. First, we measured the time taken to augment the three camera views. The results show that the domain augmentation approaches required from $3.4$s (\inpaint) to $11.4$s (\refining), while the knowledge-distilled model took only $82.0$ms (up to 138$\times$ faster).
Then, we measured the time required to execute a test scenario to evaluate the overall testing overhead introduced by the augmented domains. To ensure that the duration of the test was not influenced by external factors, such as vehicles blocking intersections, pedestrian crossings, or red lights, we used a controlled scenario without such elements.
With the diffusion models, the test runtime increased significantly, with observed test runs of $150.7$ minutes (\instruct), $358.5$ minutes (\inpaint), and $474.5$ minutes (\refining) for a single test run. In contrast, using the lighter knowledge-distilled model, the runtime increased by only $19.9$ minutes, a $20.0\%$ increase over the baseline without augmentation. We believe that the overhead is higher than single-image ADS due to the higher number of images to be augmented (three) and their larger size (600$\times$800 height$\times$width).

\begin{tcolorbox}[boxrule=0pt,frame hidden,sharp corners,enhanced,borderline north={1pt}{0pt}{black},borderline south={1pt}{0pt}{black},boxsep=2pt,left=2pt,right=2pt,top=2.5pt,bottom=2pt]
\textbf{RQ\textsubscript{4}}: \textit{
Our approach generalizes to complex urban driving environments in CARLA and InterFuser, a multi-modal ADS. We discovered 25 red light infractions and 3 vehicle collisions, which were \textit{not detected} in the original simulator. The knowledge-distilled model reduced the augmentation time to $82$ms, resulting in an overhead of only $20.0\%$.}
\end{tcolorbox}

\subsection{Threats to Validity}\label{sec:ttv}

\noindent\textbf{Internal validity.}
We utilized widely used model architectures and simulators from the literature. 
The selection of the semantic validity threshold poses another potential threat. In this study, we adopted a conservative threshold to minimize the inclusion of semantically invalid images. 
We also assumed that domain augmentations preserve driving action labels. Although similar work has made this assumption~\cite{DBLP:conf/kbse/ZhangZZ0K18,DBLP:conf/icml/LiPZL21}, we explicitly excluded domains that are unlikely to maintain the integrity of the label, such as snow or rain, as these conditions can alter the dynamics and driving style of the vehicle due to changes in friction or traction.

\noindent\textbf{External validity.}
We considered limited instances of diffusion models. To address this threat, we selected state-of-the-art diffusion models of different types that consistently improved the simulator across ODDs. 

\noindent\textbf{Reproducibility.}
To support reproducibility, all of our data, including the code of the diffusion models and our enhanced simulator, are available in our replication package~\cite{replication-package}.
\section{Related Work}\label{sec:related}

\subsection{Test Generation for Autonomous Driving}\label{sec:related:sim:scenario-based}

Existing work leverages the ability of driving simulators to create diverse driving scenes for \textit{scenario-based testing} of ADS~\cite{DBLP:journals/tosem/TangZZZGLGLMXL23, DBLP:conf/sigsoft/LouDZZ022, DBLP:journals/corr/abs-2112-00964}. Generated scenarios~\cite{DBLP:conf/icalt/JullienMVW09} include a wide range of driving conditions, such as sudden lane changes, adverse weather, or interactions with other vehicles and pedestrians~\cite{DBLP:conf/icalt/JullienMVW09}.
Majumdar et al.~\cite{DBLP:conf/fase/MajumdarMPSZ21} propose Paracosm, a tool that allows users to programmatically define complex driving scenarios.
Woodlief et al.~\cite{s3c} propose a framework that abstracts sensor inputs to coverage domains that account for the spatial semantics of a scene.
A new technique called Instance Space Analysis was recently proposed to identify the significant features of test scenarios that affect the ability to reveal the unsafe behavior of ADS~\cite{isa}. 

All of these test generators operate within a confined range of predefined ODD scenarios, including specific weather conditions, background locations, and times of day, to maximize the number of failures \textit{within} these predefined scenarios.
Our approach seeks to considerably broaden the range of ODD conditions \textit{beyond} those currently available. Our methodology is complementary and can be integrated with existing test generators to enhance their effectiveness without modifications.

\subsection{Offline Testing with Generative AI}
\label{sec:related:sim:gen-ai-testing}

Approaches based on GenAI focus on augmenting existing image datasets by introducing variations like adverse weather or other visual elements~\cite{DBLP:conf/nir/OstankovichYRG17, DBLP:conf/dsa/GaoWWWJ21}.
For example, Zhang et al.~\cite{DBLP:conf/kbse/ZhangZZ0K18} propose DeepRoad, a solution that utilizes UNIT~\cite{DBLP:conf/nips/LiuBK17} to generate test images by altering the weather from sunny to foggy or snowy. Pan et al.~\cite{DBLP:conf/qrs/PanAF21} present a method that leverages CycleGAN~\cite{DBLP:conf/iccv/ZhuPIE17} combined with techniques to synthesize different fog levels with controllable intensity and direction in driving images. Li et al.~\cite{DBLP:conf/icml/LiPZL21} propose TACTICS, an ADS testing framework that uses search-based strategies to identify critical environmental conditions and employs MUNIT~\cite{DBLP:conf/eccv/HuangLBK18} to reproduce these conditions in existing driving images. 
Attaoui et al.~\cite{attaoui2024searchbaseddnntestingretraining} combine GenAI and search-based testing to test the semantic segmentation module of an ADS.
Other approaches augment existing test images with diffusion models~\cite{DBLP:journals/csur/YangZSHXZZCY24}. Zhao et al.~\cite{DBLP:journals/corr/abs-2404-09111} exploit semantic segmentation maps and a conditional generative model, ControlNet~\cite{DBLP:conf/iccv/ZhangRA23}, to generate high-quality synthetic images. Xu et al.~\cite{DBLP:journals/jstsp/XuNCZKXMH23} employed a fine-tuned Stable Diffusion~\cite{DBLP:conf/cvpr/RombachBLEO22} to create controllable traffic signs.

Although these approaches assess the behavior of the ADS, they target \textit{model-level testing} and measure the discrepancy between predicted and ground truth values~\cite{2023-Stocco-EMSE}.
In contrast, we focus on \textit{system-level testing}. Our application of GenAI as a rendering engine within a physics-based simulator constitutes a novel contribution to the state-of-the-art in ADS testing. 

\subsection{Data-driven Simulation}

Neural simulators~\cite{DBLP:journals/corr/abs-2405-03520} 
consist of data-driven approaches in which GenAI is used to produce a continuous stream of driving images.
Unlike traditional simulators~\cite{DBLP:conf/corl/DosovitskiyRCLK17}, which rely on game-based 3D rendering and physics models, neural simulators employ a learnable world model~\cite{DBLP:journals/corr/abs-2403-02622} to represent the environment of the ADS and target novel view synthesis (e.g., bird-eye's view)~\cite{DBLP:journals/csur/XiaX24}.
For example, DriveGAN~\cite{DBLP:conf/cvpr/KimP0F21} utilizes GANs to create driving scenarios with controllable weather conditions, traffic objects, and backgrounds. 
DriveDreamer~\cite{DBLP:journals/corr/abs-2309-09777} and GAIA-1~\cite{DBLP:journals/corr/abs-2309-17080} employ diffusion models to generate real-world driving scenarios. 
UniSim~\cite{DBLP:conf/cvpr/00030WMMYU23} is a neural closed-loop sensor simulator that transforms a single recorded log from an ADS into a realistic multi-sensor simulation.

While neural simulators offer an improvement in generating novel and realistic training data, their lack of a physical representation limits their applicability for testing. This deficiency can produce inaccurate failure simulations (e.g., collisions), resulting in false positives. Thus, neural simulators are not the best choice for testing ADS systems at the system level.

To address this limitation, our approach integrates neural rendering and GenAI techniques with a physics simulator. This combination enables effective testing with precise failure determination while expanding the ODD conditions for testing.
\section{Conclusions and Future Work}\label{sec:conclusion}

We have generated new ODD scenarios for ADS testing using diffusion models and instruction editing operations. We addressed the validity of the augmented images by creating an automated semantic validator, which was found to be extremely accurate in a human study, with as few as 3\% invalid images incorrectly regarded as valid.
We have considered the realism of the augmented images by conducting a human study that indicated the \refining strategy as the technique that generates the most realistic images.
We have reduced the simulation overhead by introducing a CycleGAN model that takes advantage of knowledge distillation. Most importantly, we have shown that our approach can expose ADS failures even in domains close to the training domain and in track sectors that were deemed error-free when considering only simulator-generated test scenarios.


\bibliographystyle{IEEEtran}  
\balance
\bibliography{bibl}

\end{document}